\documentclass[11pt,dvips]{article}

\usepackage{epsfig,times} 
\usepackage{picinpar}
\usepackage{wrapfig}
\usepackage{floatflt}
\makeatletter
\renewcommand{\section}{\@startsection{section}{1}{0in}
	{0.4\baselineskip}{0.1\baselineskip}{\Large\bf}}
\renewcommand{\subsection}{\@startsection{subsection}{2}{0in}
	{0.25\baselineskip}{-\baselineskip}{\large\bf}}
\renewcommand{\subsubsection}{\@startsection{subsubsection}{3}{0in}
	{0.1\baselineskip}{-\baselineskip}{\normalsize\bf}}
\makeatother
\begin{document}

%
%  Session and Paper Code:
\makeatletter\newcommand{\ps@icrc}{
\renewcommand{\@oddhead}{\slshape{HE.6.3.15}\hfil}}
\makeatother\thispagestyle{icrc}
%
%  ***INSTRUCTIONS:***  Replace `OG 9.9.9' in the command argument below
%			with your assigned session and paper code:
\markright{HE.6.3.15}
\begin{center}
%
%  ***INSTRUCTIONS:***  Replace `Instructions for Preparation of Manuscript'
%			with your paper's title:
{\LARGE \bf Initial Results from a Search for Lunar Radio Emission from \
Interactions of ${\bf \ge 10^{19}}$ eV Neutrinos and Cosmic Rays
\footnote{\it To be published in the Proceedings of the 26th International Cosmic Ray Conference, Salt Lake City, Utah, August 1999}
}
\end{center}

%  Author List:
\begin{center}
%
%  ***INSTRUCTIONS:***  Replace authors and addresses below with your own:
%
{\bf P.~W. Gorham, K.~M. Liewer, and C.~J. Naudet}\\
{\it Jet Propulsion Laboratory, 4800 Oak Grove Dr. Pasadena, CA, 91109, USA}
\end{center}

%\begin{center}
%\end{center}

%  Abstract:
\begin{center}
{\large \bf Abstract\\}
\end{center}
\vspace{-0.5ex}
%
%  ***INSTRUCTIONS:***  Replace text below with your own abstract:
%
Using the NASA Goldstone 70m antenna DSS 14 both singly and in 
coincidence with the 34 m antenna DSS 13 (21.7 km to the southeast), 
we have acquired approximately 12 hrs of livetime in
a search for predicted pulsed radio emission from extremely--high energy 
cascades induced by neutrinos or cosmic rays in the lunar regolith. 
In about 4 hrs of single antenna observations, we reduced our
sensitivity to impulsive terrestrial interference to a negligible level
by use of a veto afforded by the unique capability of DSS 14.
In the 8 hrs of dual--antenna observations, terrestrial interference is
eliminated as a background. In both observing modes the thermal noise floor
limits the sensitivity. We detected no events above statistical background. 
We report here initial limits based on these data which 
begin to constrain several predictions of the flux of EHE neutrinos.
%

%  Leave this line skip in place:
\vspace{1ex}
\section{Introduction}
Detection of extremely high energy (EHE) neutrinos is expected to play 
an important diagnostic role in understanding the radiative kinematics of a number of cosmologically
significant phenomena. Active Galactic Nuclei (AGN) may be copious producers
of such neutrinos via hadronic interactions of particles energized
by Fermi acceleration mechanisms at the
predicted accretion shock near the central black hole. Cosmic gamma-ray
bursts are expected to produce both cosmic rays and neutrinos through
rapid shock acceleration during the burst and its gamma-ray afterglow.
EHE cosmic rays themselves interact with a $\sim 10$ Mpc mean free path with
cosmic background photons to produce other EHE neutrinos. More exotic
mechanisms such as topological defects or cosmic strings will also lead to
EHE neutrino production. 

Earth-based observatories for EHE neutrinos can only
achieve adequate sensitivity by instrumenting a huge target mass. 
Optical Cherenkov detectors such as
AMANDA and the European initiatives will in the next
few years begin approaching the necessary sensitivity to detect 
cosmic neutrinos in the $10^{12-15}$ eV range. However, 
the EHE ($\geq 10^{18}$ eV) regime is a much more
difficult problem, and none of the existing instruments appear to have
capability to address it. The required volume of water or ice to detect the
extremely low flux densities at $\geq 10^{18}$ eV  is in range of
hundreds of cubic km per year of operation. 

One proposed approach to this is to use detectors in the radio regime,
sensing at large distances the emission of coherent radio Cherenkov 
radiation from the 
excess charge
that develops in a EHE particle cascade. This method was first suggested by 
Askaryan (1962, 1965), and elaborated on by Dagkesamanskii \&
Zheleznykh (1989), whose work motivated one prior EHE neutrino search
looking for radio emission from cascades in the lunar regolith
which reported a null result
in 10 hours of livetime (Hankins, Ekers, \& O'Sullivan 1996).

\subsection{Emission Mechanisms \& LINAC Results.} 

Recent work on modeling of coherent Cherenkov
radio emission mechanisms in ice (Zas, Halzen \& Stanev 1992) 
has provided detailed simulations the emission of radio Cherenkov
radiation for cascade energies up
to $\sim 10^{16}$ eV. This initial work was later extended by 
Alvarez-Mu\~niz \& Zas (1997, 1998) in energy up to $10^{20}$ eV,
including corrections as the Landau--Pomeranchuk--Migdal (LPM) effect.
Related work by Alvarez--Mu\~niz \& Zas (1996) considered the effects
of the lunar regolith density and surface geometry on the
propagation of the pulsed emission.

The conclusions of all of these modeling efforts have only
strengthened the conclusion that lunar
cascades should produce measurable pulsed emission at cm wavelengths.
In fact, the radio emission is expected
to peak at $\sim 1-2$ GHz, and extend up to $\sim 10$ GHz before
loss of coherence begins to strongly attenuate the signal. Thus the
emission is well-matched to the receiving capabilities of many radio
telescopes in the 2--20 cm-wavelength region.
\begin{figwindow}[0,r,%
{\mbox{\epsfig{file=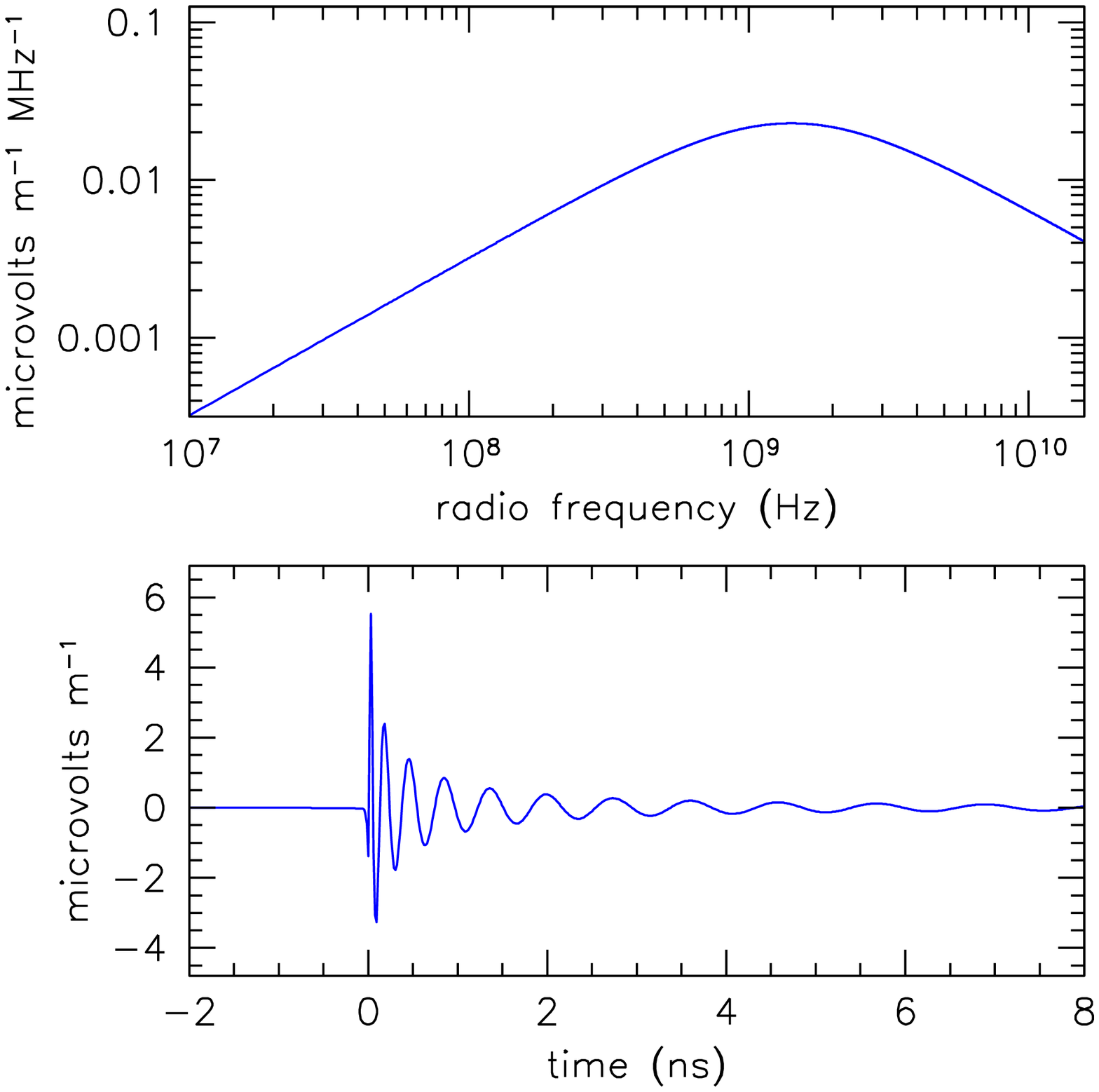,width=4.0in}}},%
{Expected spectrum and shape for a $10^{20}$ eV lunar cascade radio pulse,
including typical ionospheric dispersion. Based on models of
Alvarez-Mu\~niz \& Zas 1996, 1997, 1998.}]
Figure 1 plots the expected RF spectrum (upper) and pulse shape
(lower) from a $10^{20}$ eV lunar
cascade as received at Earth, including the effects of the ionosphere
(total electron content $10^{17}$ cm$^{-2}$). At high frequencies
($\geq 2$ GHz) the effects of the ionosphere are negligible and the
pulse duration is of order 1 ns.

There are now laboratory measurements of coherent Cherenkov and
transition radiation from electron bunches in 
electron linear accelerators
(Ohkuma et al 1991; Shibata et al 1991;
Takahashi et al 1994). These experiments confirm the
quadratic dependence of output power on electron number, and 
also verify a number of important predictions relevant to the
emission process, such as: formation zone effects, finite track--length
effects, and interference behavior related to the coherence of
the emitted waves. Thus we can begin to have some confidence that
predictions of lunar cascade emission should be taken seriously.
\end{figwindow}

\section{Goldstone Experiment}

To investigate the possibility of detecting pulsed lunar cascade
emission, we have begun a series of single and dual antenna measurements
using the NASA Goldstone Deep Space Station (DSS).
The primary antenna is DSS 14, a 70 m diameter 
Cassegrain--design antenna equipped with 1.6-1.7 GHz (L band)
single polarization, and
2.2-2.3 GHz (S band) dual polarization receivers, among others.
For our dual antenna observations, DSS 14 is paired with DSS 13, 
a 34 m antenna 21.7 km southeast
of DSS 14, connected by a high bandwidth analog optical fiber link.
The received electric field is converted to a band-limited
voltage at the receiver, then mixed with a local oscillator
to produce an intermediate frequency (IF) of 320 MHz, about which
the bandpass of interest appears on the lower sideband.

Our experimental setup for single--antenna measurements involved
a trigger system which required a $\geq 5$ ns time overlap
of fixed--width (10-20 ns) logic pulses generated directly from the 
S--band left (LCP) and
right (RCP) circular polarization IF bandpass voltages
and a fast discriminator. Since Cherenkov emission is 100\% linearly
polarized, a pulse of Cherenkov radiation will produce an
equal amplitude voltage pulse in both LCP and RCP. This 
type of coincidence is very effective at rejecting interference from
microwave telemetry systems (such as those on earth-orbiting
satellites) since these are almost universally circularly polarized.

However, such a coincidence is sensitive to impulsive interference,
such as that generated by any short-path high-voltage arcing phenomena,
which are also typically linearly polarized. Thus an enhancement to the
trigger was developed which utilized a second receiver and the
L-band feed horn. When DSS 14 is used at S-band, The L-band
feed views a region of the sky about $0.5^{\circ}$ away. Tests of
the off-axis L-band response to an on--axis S-band source showed it to be
negligible. Local terrestrial impulsive interference couples to the 
feed horns either by ``flashing'' across the sky, or scattering directly
from antenna structure (such as the subreflector). Such interference 
also tends to be steep-spectrum, falling off quickly at higher 
frequencies. 

We found in practice that, by using the L-band 
signal as a veto, we could virtually eliminate any triggers from
terrestrial interference. This was tested several times at DSS 14, including once under extreme conditions when the trigger rate (for $>4.5 \sigma$ pulses
in both polarizations) was 1.5 kHz without the veto. With the
veto turned on, the rate of interference triggers decreased to $\leq 0.01$ Hz,
a factor of $\geq 10^{5}$ rejection. Such conditions are
atypical (average rates of interference triggers are one every few minutes
or less), but demonstrate clearly the effectiveness of this method. 

Once a trigger is generated, the IF voltage time series for each of
the signals present: S-band LCP, S-band RCP, and L-band LCP (veto), are
sampled at 1 GHz using a Tektronix TDS784 digital oscilloscope,
then read out to computer. Post--analysis of the recorded triggers
requires that they are:
(1) band-limited in width, consistent with expectations that the
cascade pulses are $\sim 1$ns in duration;
(2) not significantly offset in time, or greatly different in
amplitude in RCP and LCP; 
(3) without any significant counterpart in the L-band recorded
signal, indicating that they are not likely to be terrestrial
in origin;
(4) of sufficient amplitude in each polarization that they are 
statistically very unlikely from random thermal noise fluctuations.

In the case of the dual antenna system, an additional criterion is
included with those above: (5) A band-limited pulse must appear in the DSS 13
IF data coincident in the time window that is consistent with the
geometric delay to the surface of the Moon viewed by the two antennas.
In practice the width of this window must be of order $B\Theta_mc^{-1}\approx
\pm 290$ns (where $B$ is the baseline and $\Theta_m$ is the 
angular size of the Moon) to account for cascades 
at any point on the Moon's surface.

\section{Results}

Our hardware trigger was as loose as the data collection system could
tolerate, giving trigger rates of 0.1 Hz for the single-- and 0.01 Hz for
the dual--antenna operation. Since the great majority of these triggers 
were random thermal noise coincident pulses of the two polarizations
our post-processing of the data involved setting a threshold which was
well beyond any possibility of random noise coincidence.
Thus we required the signals from the 70m antenna (DSS 14)
to exceed the $6 \sigma$ level in both LCP and RCP, along with an
additional requirement of $\ge 4 \sigma$ from the DSS 13 signal. These
requirements can be transformed into an energy threshold for the
system, based on the cascade simulations noted above and the standard
equation for radio antenna sensitivity:
$\Delta S ~=~ 2kT_{sys} A_{eff}^{-1} (\Delta t \Delta \nu )^{-0.5}$
W m$^{-2}$ Hz$^{-1}$. 
The $1\sigma$ thermal noise level for our system is
$\sim 400$ Jy, giving a post-analysis $6\sigma$ threshold of
2400 Jy. This in turn corresponds to the flux density expected from
a $\sim 1 \times 10^{19}$ eV cascade. Given that virtually all cascades
at these energies are expected to be primarily hadronic in nature,
the mean of the Bjorken y-distribution implies a mean neutrino energy of
a factor of 5 greater than this. Thus we estimate our neutrino energy
threshold to be $5 \times 10^{19}$ eV, with the peak sensitivity
(weighted by effective volume and an $E^{-2}$ spectrum) at an
energy about a factor of 4--6 higher than this.

We observed the Moon for 4.8 hr on the Moon center, 
5.6 hrs at the midpoint from center to limb, and
1.5 hrs at the limb, for a total of 11.9 hrs.
Of the $\sim 300$ (in 8 hrs) dual antenna and $\sim 1100$ (in 4 hrs) 
single antenna
triggers, none matched all of the criteria above
\footnote{In both the single-- and
dual--antenna case, the vast majority of the triggers are simple
thermal noise coincidences near the hardware threshold of $4.5 \sigma$
in each polarization. The remaining triggers are obvious RF
interference where the veto pulse level fell below the more
conservative ($\geq 5 \sigma$) veto threshold, but is still clearly
present in the sampled L-band IF data.}

\subsection{Limits on EHE neutrinos.}

   Figure 2 plots the predicted fluxes of EHE neutrinos from a number of
models including AGN production (Mannheim 1996, M)
gamma-ray bursts
(Bahcall \& Waxman 1999, BW), EHE cosmic-ray interactions (Hill \& Schramm
1985, HS),
and topological defects (TDs; Yoshida et al 1997, YTD; Bhattacharjee et al
1992, BTD; Sigl et al 1996, STD). 
Also plotted are limits from about 70 days of Fly's Eye livetime 
(Baltrusaitas et al 1985).
Our initial 90\% CL limit is shown plotted 
with inverted triangles.\footnote{
The behavior of our differential limit was determined by
semi--analytic modeling of the effects of the antenna beam response,
the interaction geometry, and RF attenuation of the regolith.}
\begin{figwindow}[2,r,%
{\mbox{\epsfig{file=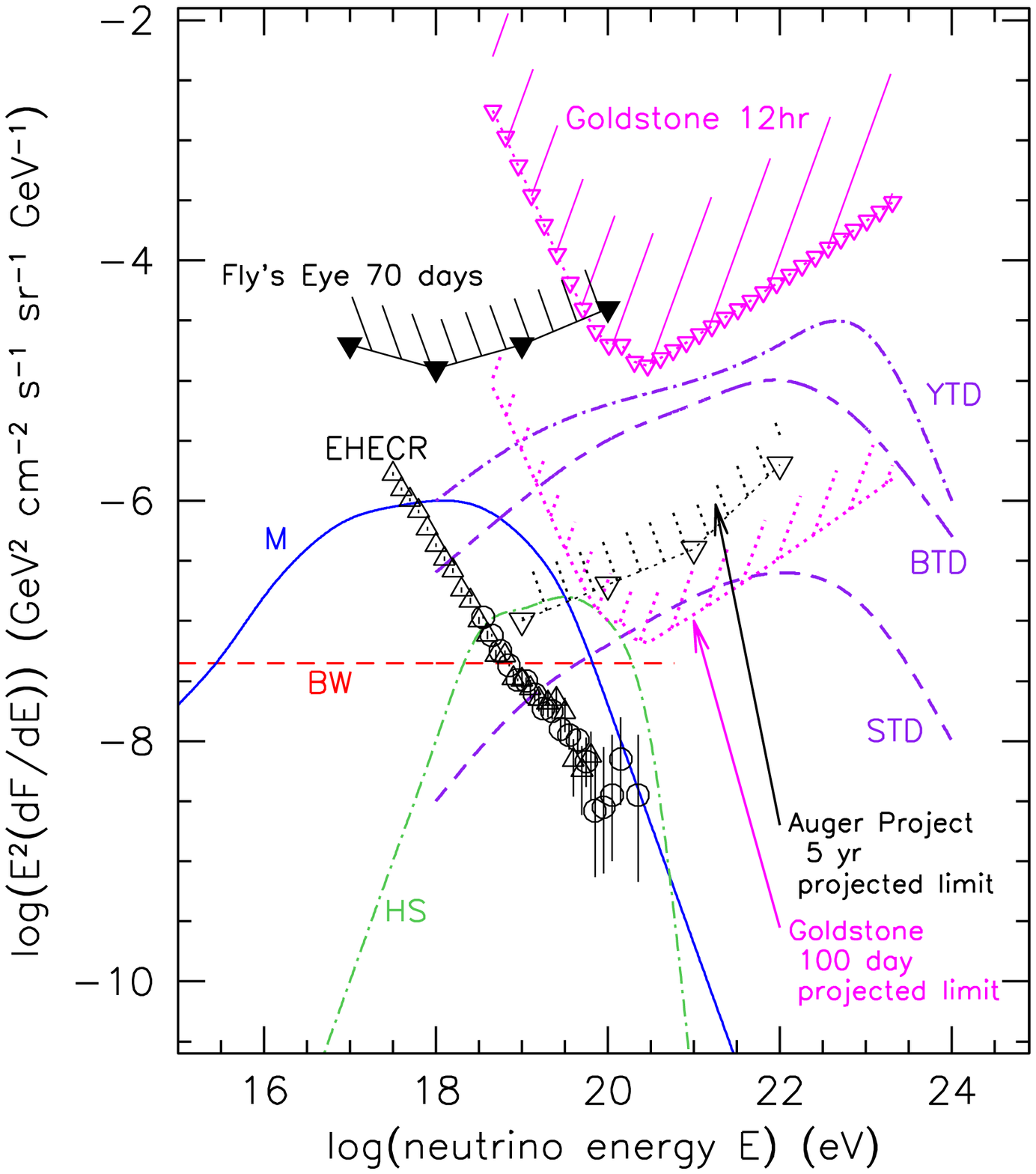,width=4.0in}}},%
{Predictions \& limits for EHE neutrinos (details in text).}]
Our limits just begin to constrain the
TD model of Yoshida et al (1997) for which 
we expected a total of order 1--2 events.
Also plotted in Fig. 2 are projected 5 yr livetime limits for the Auger Project
(Capelle et al 1998), and projected Goldstone limits for 100 days of
livetime, which would take several years to accumulate, 
but would test several of the present models.
Fig. 2 also shows the EHE cosmic ray (EHECR) spectrum on the same scale as the 
neutrino predictions and limits. Alvarez--Mu\~niz \& Zas (1996) estimated
that the EHECR would dominate the event rates for EHE lunar radio
searches. For their predicted fluxes
we should have seen 3--4 EHECR--induced events in our data,
but saw none. However, we conclude that this prediction
is likely to be an overestimate, for two
reasons not considered by Alvarez--Mu\~niz \& Zas: 
(1) The geometry of EHECR cascade emission causes total
internal reflection in the regolith to first order; (2) Formation
zone effects suppress emission from cascades that ``skim''
the regolith surface. We conclude that only narrow ridglines or
similar features will be efficient at EHECR
radio emission. Expected rates are far below our limit.
\end{figwindow}
{\bf Acknowledgements.} We thank M. Klein, T. Kuiper, G. Resch, L.
Teitelbaum, \& the Goldstone Operational Staff
for their support of this work, which was performed 
by the Jet Propulsion Laboratory, Calif. Inst. of Technology, under 
contract with the National Aeronautics and Space Administration.
%
%
%  References: (DO NOT ALTER NEXT 4 LINES)
\vspace{1ex}
\begin{center}
{\Large\bf References}
\end{center}
%
%  ***INSTRUCTIONS:***  Enter your references alphabetically following the format
%			of the example citations below.
Alvarez--Mu\~niz, J., \& Zas, E., 1997, Phys. Lett. B, 411, 218; also 
1998, LANL preprint astro-ph/9806098. \\
Alvarez--Mu\~niz, J., \& Zas, E., 1996, Proc. XXVth ICRC, ed. M.S. Potgeiter,
et al. vol. 7, 309.\\
Askaryan, G.A.,1962, JETP 14, 441; also 1965, JETP 21, 658. \\
Bahcall, J.N., and Waxman, E., 1999, LANL preprint astro-ph/9902383 .\\
Baltrusaitas, R.M., Cassiday, G.L., Elbert, J.W., et al 1985, Phys Rev D 31, 2192.\\
Bhattacharjee, P., Hill, C.T., \& Schramm, D.N, 1992 PRL 69, 567.\\
Capelle, K.S., Cronin, J.W., Parente, G., \& Zas, E., 1998, Astropart. Phys.
8, 321.\\
Dagkesamanskii, R.D., \& Zheleznyk, I.M., 1989, JETP 50, 233. \\
Hankins, T.H., Ekers, R.D. \& O'Sullivan, J.D. 1996, MNRAS 283, 1027.\\
Hill, C.T.,  \& Schramm, D.N., 1985, Phys Rev D 31, 564. \\
Mannheim, K., 1996, Astropart. Phys 3, 295.\\
Ohkuma, J., Okuda, S., \& Tsumori, K., 1991, PRL 66, 1967. \\
Shibata, Y. , Ishi, K., Takahashi, T., et al, 1991, Phys Rev A 44, 3449. \\
Takahashi, T, Kanai, T., Shibata, Y., et al, 1994, Phys Rev E 50, 4041.\\
Yoshida, S., Dai, H., Jui, C.C.H., \& Sommers, P., 1997, ApJ 479, 547.\\
Zas, E., Halzen, F., \& Stanev, T., 1992, Phys Rev D 45, 362. \\

\end{document}